\documentclass[aps,prl,superscriptaddress,twocolumn,longbibliography,floatfix]{revtex4-2}

\usepackage{braket}
\usepackage{bm}
\renewcommand\vec{\bm}
\usepackage{float}
\usepackage{graphicx}
\usepackage{amsmath}    
\usepackage{comment}
\usepackage{units}
\usepackage{amsthm}
\usepackage{amssymb}
\usepackage{graphicx}
\usepackage{color}
\usepackage{bbold}
\usepackage[colorlinks=true, urlcolor=blue, citecolor=blue,linkcolor=blue,citebordercolor={1 0 0},linkbordercolor={0 0 1}]{hyperref}
\usepackage{graphicx}
\graphicspath{{./images/},{./imagesAppendix/}}

\newcommand{\mel}[3]{\langle #1 \vert #2 \vert #3\rangle}
\newcommand{\tr}[1]{\text{Tr}\left(#1\right)}

\renewcommand{\eqref}[1]{Eq.~(\ref{#1})} % Reference to equation
\newcommand{\figref}[1]{Fig.~\ref{#1}} % Reference to figure
\theoremstyle{plain}
\newtheorem{thm}{\protect\theoremname}
\theoremstyle{plain}

\ifx\proof\undefined
\newenvironment{proof}[1][\protect\proofname]{\par
	\normalfont\topsep6\p@\@plus6\p@\relax
	\trivlist
	\itemindent\parindent
	\item[\hskip\labelsep\scshape #1]\ignorespaces
}{%
	\endtrivlist\@endpefalse
}
\providecommand{\proofname}{Proof}
\fi
\theoremstyle{plain}

\theoremstyle{remark}

\makeatother
\newcommand{\idg}[1]{{\bfseries #1)}}

\providecommand{\factname}{Fact}
\providecommand{\theoremname}{Theorem}
\providecommand{\claimname}{Claim}
\providecommand{\lemmaname}{Lemma}
\providecommand{\definitionname}{Definition}

\theoremstyle{definition}
\newtheorem{defn}[thm]{\protect\definitionname}

\usepackage{amssymb}
\usepackage{color}
\usepackage{color}
\definecolor{KB}{rgb}{0.4,0.3,0.9}

\begin{document}

\title{Convex optimization for non-equilibrium steady states on a hybrid quantum processor}
% \thanks{}%
\author{Jonathan Wei Zhong Lau}
\thanks{Equal contribution}
\affiliation{Centre for Quantum Technologies, National University of Singapore, 3 Science Drive 2, Singapore 117543}
\author{Kian Hwee Lim}	
\thanks{Equal contribution}
\affiliation{Centre for Quantum Technologies, National University of Singapore, 3 Science Drive 2, Singapore 117543}
\author{Kishor Bharti}
\affiliation{Joint Center for Quantum Information and Computer Science and Joint Quantum Institute, NIST/University of Maryland, College Park, Maryland 20742, USA}
\affiliation{Institute of High Performance Computing (IHPC), Agency for Science Technology and Research (A*STAR), 1 Fusionopolis Way, \#16-16 Connexis, Singapore 138632, Republic of Singapore}

\author{Leong-Chuan Kwek}
\affiliation{Centre for Quantum Technologies, National University of Singapore, 3 Science Drive 2, Singapore 117543}
\affiliation{MajuLab, CNRS-UNS-NUS-NTU International Joint Research Unit, UMI 3654, Singapore}
\affiliation{National Institute of Education,
Nanyang Technological University, 1 Nanyang Walk, Singapore 637616}
\affiliation{School of Electrical and Electronic Engineering
Block S2.1, 50 Nanyang Avenue, 
Singapore 639798 }
\author{Sai Vinjanampathy}
\affiliation{Centre for Quantum Technologies, National University of Singapore, 3 Science Drive 2, Singapore 117543}
\affiliation{Department of Physics, Indian Institute of Technology-Bombay, Powai, Mumbai 400076, India}
\email{sai@phy.iitb.ac.in}
\affiliation{Centre of Excellence in Quantum Information, Computation, Science and Technology, Indian Institute of Technology Bombay, Powai, Mumbai 400076, India.}
\begin{abstract}
    Finding the transient and steady state properties of open quantum systems is a central problem in various fields of quantum technologies. Here, we present a quantum-assisted algorithm to determine the steady states of open system dynamics. By reformulating the problem of finding the fixed point of Lindblad dynamics as a feasibility semidefinite program, we bypass several well known issues with variational quantum approaches to solving for steady states. We demonstrate that our hybrid approach allows us to estimate the steady states of higher dimensional open quantum systems and discuss how our method can find multiple steady states for systems with symmetries.

\end{abstract}

 \maketitle
\paragraph{Introduction.---}

Understanding open system evolution is central to modern
quantum technologies such as computing, thermodynamics~\cite{schwarz2018nonequilibrium,ikeda2020general,fraenkel2021entanglement}, chemistry~\cite{raeber2020non}, and quantum
transport~\cite{manzano2018harnessing}.  Since such evolution maps initial
quantum states to future states, both transient and steady state properties
are available in the structure of the evolution operator.  Sparing few
analytically tractable systems, generic open system evolution has to be
solved numerically to understand both transient and steady state dynamics of
the system.  Such classical simulation techniques are limited due to the
exponential growth of Hilbert space.  Some specific sampling problems can be
simulated
classically~\cite{foulkes2001quantum,yan2018interacting,nagy2018driven,nagy2019variational}
and tensor networks can be deployed for scenarios with limited entanglement
growth~\cite{zwolak2004mixed,verstraete2004matrix,orus2008infinite,cui2015variational,werner2016positive,gangat2017steady,kshetrimayum2017simple,verstraete2008matrix}.
For generic open system evolution by contrast, such a classical simulation is
limited to few dozen qubits in the presence of symmetries.  Usually, such
problems are either simplified by the presence of strong local dissipators
which reduce the amount of entanglement generated or by low dimensionality of
the problem.  Outside of these special cases, the issue of generic open
system evolution has remained unsolved.

The advent of small quantum computers heralds a new variety of solutions to
the problem of determining the transient and steady state solutions to such
open system evolution.  One strategy involves implementing open system
evolution on an intermediate scale quantum computer and tomographically
measuring the quantum state at various times~\cite{su2020quantum}.  An
equivalent method for completely positive maps would be to quantum simulate and measure the
Choi matrix associated with the open system
evolution~\cite{hu2020quantum,schlimgen2021quantum,liu2022solving}.  These
tomographic methods require exponentially large number of measurements and
hence are practically infeasible.  Another group of closely-related
strategies involves first implementing $\mathcal{L}$, the Liouville
superoperator associated with the open system evolution, on a quantum
computer.  After implementing $\mathcal{L}$ on a quantum computer, the
different strategies to find the non equilibrium steady states (NESS) include methods like a combination of
Trotterisation and imaginary time evolution using
$\mathcal{L}$~\cite{kamakari2022digital}, quantum phase estimation on
$\mathcal{L}$~\cite{ramusat2021quantum}, and variational quantum algorithms (VQAs)
to find the kernel of $\mathcal{L}^\dag \mathcal{L}$
\cite{yoshioka2020variational}.  These different but related strategies have
their own individual drawbacks.  Trotterisation and phase estimation
approaches are known to be infeasible on our current quantum devices with
short coherence times, and the variational optimisation approaches suffers
from the difficulty of optimising over a non-convex
space~\cite{anschuetz2019variational,you2021exponentially,bittel2021training}.
Lastly, all of these methods that rely on the superoperator representation
$\mathcal{L}$ of the open system evolution suffer from the large
dimensionality of the Liouville space.

In this paper, we propose a hybrid algorithm for the determination of NESS.
Through our approach, the steady state problem can be recast as
solving a feasibility semidefinite program
(SDP)~\cite{vandenberghe1996semidefinite,boyd2004convex,wolkowicz2012handbook}.
We show that such an approach to find the NESS is viable on a NISQ device.  Our first contribution is to restate the NESS
problem as a feasibility SDP, which is an SDP where the goal is to find a
feasible solution satisfying the positive semidefinite and linear
constraints~\cite{vandenberghe1996semidefinite,boyd2004convex,wolkowicz2012handbook}.
Our second contribution is that we do not use a variational quantum
state/circuit as the
ansatz~\cite{yoshioka2020variational,liu2021variational,cerezo2021variational,bharti2021noisy}.
By doing so, we bypass the problems~\cite{bharti2021nisq,
haug2020generalized,lau2021nisq,CQFF} associated with
training variational quantum algorithms with their non-convex landscape,
which is known to be non-deterministic polynomial-time (NP)
hard~\cite{bittel2021training,anschuetz2019variational,you2021exponentially}.
We show that our algorithm naturally enforces positivity constraint of a
physical density matrix and provides methods to enforce additional
constraints systematically while retaining the advantages of
quantum-assisted
methods~\cite{bharti2020quantum,bharti2021iterative,haug2020generalized,
lau2021noisy,CQFF}, like providing a method to systematically gain a more expressible, problem aware ansatz. 
\paragraph{Non-equilibrium steady states.---}
Open system dynamics under Born, Markov and secular approximations are often
described by a time-local master equation given by $\dot{\rho}=L[\rho]$ where
\begin{align}
  L[\rho] &= -i [H,\rho] + \sum_n\gamma_n \left(A_n \rho A_n^\dag
  -\frac{1}{2}\{A_n^\dag A_n ,\rho \} \right). \nonumber
  %\label{eqn: masterequation_ness}
\end{align}

Such an evolution preserves conditions for valid
density matrices.  The transient and steady states of this evolution are
characterized by the spectrum of the Liouville superoperator~\cite{manzano2018harnessing}, defined by the vectorization $B\rho
C\rightarrow C^*\otimes B\ket{\rho}$.  Steady states are understood to
satisfy $L[\rho] = 0$ or equivalently $\mathcal{L}\ket{\rho}=0$, where
$\mathcal{L}$ is the Liouville superoperator that arises from the
vectorisation of $L$.  Since these steady states do not usually correspond
to a thermal equilibrium, they are referred to as non-equilibrium steady
states (NESS).  We refer to the problem of obtaining the steady state(s) of a
given Liouville evolution as the NESS problem, which is solved classically by
matrix diagonalization.  However, due to the increase in dimensionality,
diagonalization of the full spectrum is usually unfeasible.  Furthermore, the evolution of $n$-dimensional density matrices in Liouville space
are represented by $n^2\times n^2$ matrices.  This squared dimensionality
implies that numerical techniques can find the entire spectrum of only modest
open quantum systems, usually relying on Arnoldi type
methods~\cite{lanczos1950iteration,arnoldi1951principle,lehoucq1998arpack,saad2011numerical}, which become quite cumbersome for many-body systems of
moderate size. 

Hence, there is interest in understanding if quantum computers, with their
inherent dimensionality advantages in simulating quantum systems over
classical computers, can solve the NESS problem.
For NISQ devices, it was shown that the NESS problem can be mapped
to a variational problem in Liouville space \cite{yoshioka2020variational}.
The subsequent variational problem is solved by using a parameterized quantum
state or quantum circuit as the ansatz, and relies on forms of VQA.
This approach has two main concerns.
Firstly, it is unclear how to
systematically enforce the positivity constraint for the density matrix in
this approach, as the variational quantum state/quantum circuit, which is a
vector, must eventually correspond to a physical density matrix using the
vectorisation described above.  Secondly, optimizing over the set of pure
states tends to not be convex and hence difficult,
% which makes the optimization difficult,
and indeed has been shown to be NP-hard, reasons including the parameter
landscape containing exponentially many persistent local minima that are far
from the global
minimum~\cite{bittel2021training,anschuetz2019variational,you2021exponentially}
(See Supplemental Material \cite{referenceToSupplementalMaterial}).
% (See Appendix~\ref{appendix:comparison}).
Other VQA methods that do not
explicitly rely on this map to Liouville space~\cite{liu2021variational} face
similar problems.

\paragraph{Quantum Feasibility SDP Approach.---}
We circumvent the non-convex optimization problem in the Liouville space by
optimizing over the convex set of density matrices.  This allows us to
directly apply a feasibility SDP, one consequence of which is that we can now
systematically enforce the positive semidefinite condition.  A feasibility
SDP admits the following form: $\text{Find }X,\, X\in S_+^{l},\,\text{such
that } \text{Tr}(C_k X)=v_k, \,\forall k\in{1,2,\dots c}$.
Here, $\mathcal{S}_+^{l}$ represents the set of $l\times l$ symmetric
PSD matrices. This is the problem of determining if it is possible to find a matrix $X$ subject to the PSD constraint and the other given constraints.  The matrices $C_k$ belong to the set of
symmetric matrices $\mathcal{S}^l$ for $k\in \{1,2,\cdots c\}$.  The $k$-th
element of vector $v\in \mathbb{R}^c$ is denoted by $v_k$.  SDPs can be
formulated for complex-valued matrices via a cone of Hermitian positive
semidefinite matrices i.e.  $X \in \mathcal{H}_+^{l}$.  Since SDPs for real
valued matrices are a special case of SDPs for complex-valued matrices, we
will consider the latter case in this paper.  Since 
$\dot{\rho}=L[\rho]$ is linear in $\rho$, the NESS problem is a feasibility
SDP.

We consider a state ansatz of the form
 \begin{equation}
  \label{eqn: denmat ansatz}
  \rho = \sum_{i,j}\beta_{ij}\ket{\chi_i}\bra{\chi_j}.
 \end{equation}

Here, $\beta_{ij}$ are matrix elements of a positive semidefinite matrix
$\beta$, whereas $\ket{\chi_i}$ states can be from any set of quantum states.
We see that $\beta$ being positive semidefinite is both a necessary and
sufficient condition for $\rho$ to be positive semidefinite.  The condition
$\tr{\rho} = 1$ becomes $\tr{\beta E} = 1$, where $E$ is a
matrix with matrix elements $E_{ij} = \braket{\chi_i|\chi_j}$.
% The trace
% condition on $\rho$ can then be re-written as
  
With the chosen ansatz, the NESS problem becomes
  \begin{gather}\label{eqn: feasibility sdp}
      \text{Find } \beta \text{ s.t.}
      -i(D\beta E- E \beta D)\notag\\
      + \sum_n \gamma_n \left(R_n \beta R_n^\dag
      -\frac{1}{2} F_n \beta E -
      \frac{1}{2} E \beta F_n\right) = 0,\\
      \beta \succcurlyeq 0, \\
      \text{Tr}(\beta E) = 1 ,
  \end{gather}
where $\gamma_n$ are the strengths of the dissipators, $D,R,F$ are matrices
defined as $D_{ij} = \mel{\chi_i}{H}{\chi_j}$, $\left(R_n\right)_{ij} =
\mel{\chi_i}{A_n}{\chi_j}$ and $\left(F_n\right)_{ij} =
\mel{\chi_i}{A_n^\dagger A_n}{\chi_j}$.
% We have now reduced the 
This reduction of the NESS problem to a feasibility
SDP~\cite{boyd1994linear,boyd2004convex} defined over $\beta$ is motivated by
the idea that
% , the idea
% behind which is that
a judicious choice of the states $\ket{\chi_i}$ in some
problem-aware manner could possibly allow us to do an optimisation over a
smaller dimensional convex landscape (compared to $\rho$).
% This is a feasibility SDP~\cite{boyd1994linear,boyd2004convex} defined over
% $\beta$.  By reducing the problem to a SDP over $\beta$, the idea is that a
% judicious choice of the states $\ket{\chi_i}$ in some problem-aware manner
% could possibly allow us to do an optimisation over a smaller dimensional
% convex landscape (compared to $\rho$).
Furthermore, the positive semidefiniteness condition of $
\rho$. can be enforced
naturally.  We utilize CVX~\cite{grant2014cvx}, that relies on a disciplined
convex programming algorithm~\cite{grant2006disciplined,grant2008graph}.

% We also note that we can easily enforce 
We can also easily enforce additional linear constraints of the
form $\text{Tr}(\beta X) = x$, where $X$ and $x$ are arbitrary matrices and
values respectively. This feature of our scheme is absent in the existing algorithms for solving NESS on NISQ devices and is further discussed below.

The overlap values for the matrix elements of the $E,D,R,F$ matrices can be
measured on a NISQ quantum computer~\cite{mitarai2019methodology}. In general, how we choose the $\ket{\chi_i}$ states to form our ansatz
will contribute strongly to how our algorithm scales.  For a general
Hamiltonian, absent of exploitable symmetries, the size of the optimal
ansatz will grow exponentially with the size of the problem. (see Supplemental Information~\cite{referenceToSupplementalMaterial}).
% This issue is
% fundamentally about obtaining an appropriate ansatz that is expressible
% enough.  It is known that to prepare an arbitrary state on an $n$ qubit
% quantum computer, we require a circuit depth of at least
% $2^n/n$~\cite{knill1995approximation,mottonen2004quantum,vartiainen2004efficient,plesch2011quantum}.
% This is a complexity theoretic statement that cannot be bypassed by any
% quantum simulation algorithm based on parametric quantum circuits or linear
% combination of quantum states, and indicates that our algorithm shares the worst case of requiring an exponentially large ansatz to obtain perfect fidelity with other variational algorithms.  
% Thus, while in the worst case our algorithm
% requires an exponentially large ansatz to obtain perfect fidelity, this
% problem will be shared by all other variational algorithms.  
Even in the worst case where we require exponentially large numbers of
$\ket{\chi_i}$ states in our ansatz, we do not map the problem to an
equivalent one in Liouville space and avoid the aforementioned squared
dimensionality that comes from doing optimization in Liouville space.
Hence in the worst case, our method is at least quadratically better than
analogous variational algorithms. 

Unless otherwise stated, we choose cumulative $\mathcal{K}$ moment states (
$\mathbb{CS}_{K}$ states )~\cite{bharti2021iterative} which provide us
with a systematic way to generate an increasingly expressible problem aware ansatz. These states rely heavily on calculating expectation values of powers of the
Hamiltonian $\braket{\psi|H^k|\psi}$ which can be done efficiently~\cite{seki2021quantum,Guru}. They alternatively can also be easily obtained by 
calculating the expectation value of Pauli
strings~\cite{bharti2020quantum,bharti2021iterative} (see~\cite{referenceToSupplementalMaterial} for details). By using the $\mathbb{CS}_{K}$ states as an ansatz, the size of the $\beta$ matrix that will be calculated scales as $\delta^K$, where $\delta$ is the number of terms in the Hamiltonian, for small $K$. While this is typically not scalable, we emphasize that our method need not use $\mathbb{CS}_{K}$ states as its ansatz. Our main contribution is in approaching the steady state problem in terms of a SDP, and the choice of ansatz in our paper is secondary. A more efficient method of generating an ansatz can be used, if we have greater knowledge of the underlying symmetries of the system. Note that the SDP itself could also be sped up with the help of a quantum computer~\cite{brandao2022faster}.

The algorithm can hence be summarised as (a) choose a hybrid ansatz for $\rho$ using a set of chosen quantum states $\{\ket{\chi_i}\}$ (b) calculate the entries of the overlap matrices on the quantum computer, (c) we use the matrices in a SDP optimization routine run on a classical computer to obtain the approximate NESS.
  
\begin{figure}
    \centering
    \includegraphics[width=0.39\textwidth]{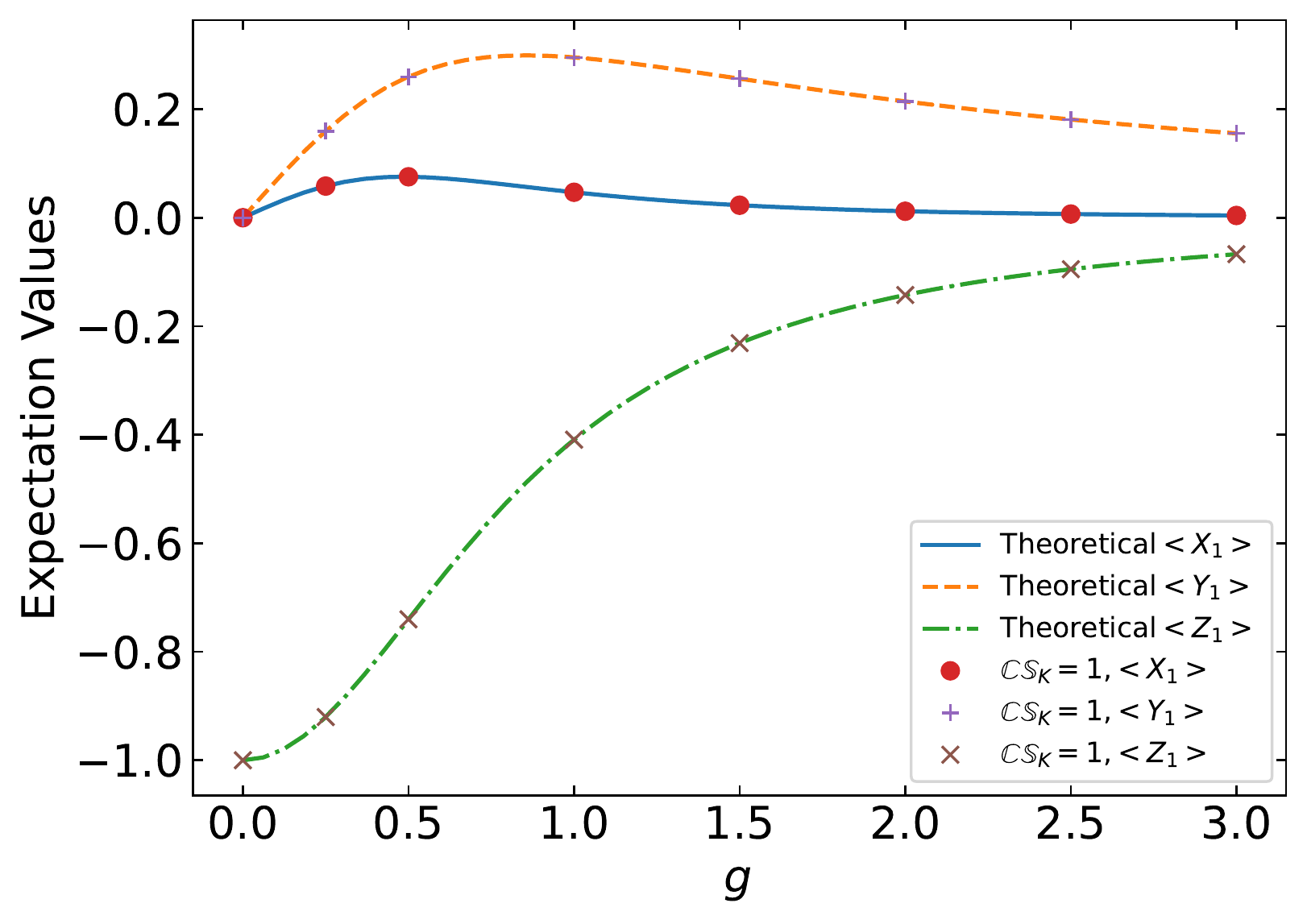}
    \caption{Expectation values for two qubit transverse field Ising model.
    $\gamma$s set at 1.  Fidelity is equal to $1$ for all values of $g$.  Our
    method gives strong agreement with the theoretical results.}
    \label{fig:2_qubit_expectation}
\end{figure}

\begin{figure}
   \centering
   \includegraphics[width=0.44\textwidth]{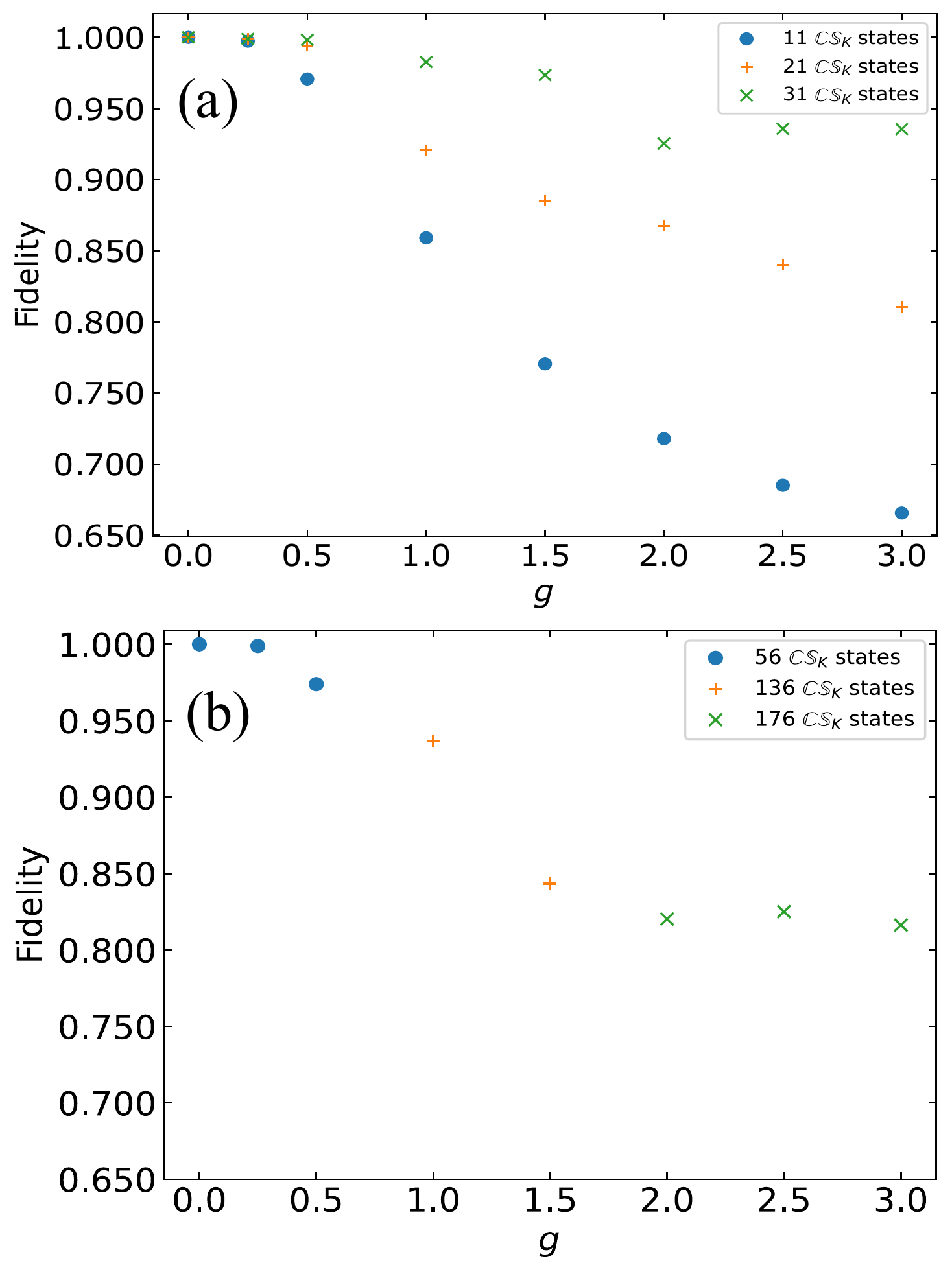}
   \caption{Results for the transverse field Ising model with local dissipators described in the main text. The corresponding fidelity value
   between the state obtained and the theoretical state, for $\mathbb{CS}_{K}$ ansatz of different ansatz sizes $K$, are compared. \idg{a} Results for 5 qubits. \idg{b} Results for 8 qubits. For larger $g$, we note that the exact NESS becomes much less sparse. To continue to obtain good fidelities in this regime, we require larger number of states in our ansatz. 
  %  Extra results for 8 qubits are given in the Supplemental Material~\cite{referenceToSupplementalMaterial}.
  }
   \label{fig:5_and_8_qubit_combined}
\end{figure}

\paragraph{Examples.---}
We demonstrate our algorithm with some examples.  Consider a
two qubit transverse field Ising model with the Hamiltonian $H_2 =
(1/2)\sigma_Z^1 \sigma_Z^2 + g \sigma_X^1 + g \sigma_X^2$, together with local
dissipators $A_1 = \sigma_Z^1$, $A_2 =
(1/2)(\sigma_X^1 - i \sigma_Y^1)$, $A_3 = \sigma_Z^2$ and $A_4 = (1/2)(\sigma_X^2 - i\sigma_Y^2)$. For all instances presented in \figref{fig:2_qubit_expectation}, our hybrid algorithm outputs a density matrix $\rho$ that is unit trace, Hermitian, positive semidefinite and that fulfils the NESS condition $\dot{\rho} = 0$.
To study the robustness of the algorithm for larger chains, in \figref{fig:5_and_8_qubit_combined} we show simulation results for the transverse field Ising model up to eight qubits. For the five qubit and eight qubit systems presented in \figref{fig:5_and_8_qubit_combined}, when increasing $K$, we choose from the $\mathbb{CS}_{K}$ ansatz, a random subset of new states, as highlighted in the Supplementary Information. A comparison with the existing NISQ approach in~\cite{yoshioka2020variational} for the eight qubit case is also given in the Supplementary Information~\cite{referenceToSupplementalMaterial}.

We note that for the model chosen, as $g$ increases, the exact NESS solution
has larger rank and is less sparse.  We find that for such situations, a
larger ansatz size is needed to obtain an approximate NESS with similar
fidelity.  We also note that the $\mathbb{CS}_{K}$ ansatz performs
efficiently when the steady states are low rank.  When this is not the case,
it is expected that any NISQ algorithm based on such ansatzes will
underperform.  Likewise, we note that another choice that significantly
influences the ansatz is the choice of initial states, where recent results
on solving the ground state problem can aid in providing useful initial
states \cite{lin2020near}.

% \begin{figure}
%     \centering
%     \subfigimg[width=0.42\textwidth]{a}{startlargesteigvec_newgraph_5_qubit_noiseless_fidelity.pdf}
%     \subfigimg[width=0.42\textwidth]{b}{diffansatzsize.pdf}
%     %\includegraphics[width=5in]{doubleslitnospin05phasedsheatmap.png}
%     \caption{\Jon{Presented here are the results for the transverse field Ising model with local dissipators, described in the main text. The corresponding fidelity value
%     between the state obtained and the theoretical state, for $\mathbb{CS}_{K}$ ansatz of different ansatz sizes $K$, are compared. \idg{a} Results for 5 qubits. \idg{b} Results for 8 qubits. For larger $g$, we note that the exact NESS becomes much less sparse. To continue to obtain good fidelities in this regime, we require larger number of states in our ansatz.} }
%     \label{fig:5_qubit}
% \end{figure}

% The results for \idg{a} indicate that increasing the cardinality of the $\mathbb{CS}_{K}$ ansatz improves results. Our results also showed that the classical regime, corresponding to high $g$, require less expressible ansatz to obtain good fidelity. The results for \idg{a} were generated with the $\mathbb{CS}_{K}$ states, for an initial state with a non-exponentially decreasing overlap with the actual steady state, by choosing one of the eigenvectors of the theoretical density matrix. The results for \idg{b} were obtained by using the $2^8$ computational basis states as the ansatz, instead of $\mathbb{CS}_{K}$ states, as a proof of principle.

\paragraph{Strong symmetries.---}
One additional complication with the NESS problem is that systems with
symmetries can exhibit multiple NESS~\cite{manzano2018harnessing}.  Our
algorithm can also be extended to certain cases where multiple NESSs are
expected.
If there is a strong symmetry in the system, then the Hilbert space can be
decomposed into the symmetry subspaces, namely
\begin{gather}
    \mathcal{H} = \bigoplus_{\alpha=1}^{n_U}\mathcal{H}_\alpha,\quad
    \mathcal{H}_\alpha =\text{Span}\left\{ \ket{\psi^{(k)}_\alpha}\right\},k\in [1,d_\alpha].
\end{gather}
Here $\ket{\psi^{(k)}_\alpha}$ are the eigenvectors of the unitary $U$ which
characterise the system's strong symmetry.  The corresponding eigenvalues are
$u_\alpha$, $\alpha \in [1,n_U]$, where $1\leq n_U \leq D$ is the number of
distinct eigenvalues of $U$, and $k \in [1,d_\alpha]$, where $d_\alpha$ is
the dimension of the subspace corresponding to the eigenvalue $u_\alpha$.
This decomposition can be extended to the operator space
$\mathcal{B}(\mathcal{H})$, through
\begin{gather}
    \mathcal{B}(\mathcal{H}) =
    \bigoplus_{\alpha=1}^{n_U}\bigoplus_{\beta=1}^{n_U}\mathcal{B}_{\alpha
    \beta},\label{eqn:invariant_subspaces}
\end{gather}

where $\mathcal{B}_{\alpha \beta} =\text{Span}\left\{
    \ket{\psi^{(n)}_\alpha}\bra{\psi^{(m)}_\beta}\right\},n\in [1,d_\alpha], m\in
    [1,d_\beta]$. Each orthogonal subspace can each contribute to the NESS solution, since each
subspace $\mathcal{B}_{\alpha\beta}$ can have a solution $\rho_{\alpha\beta}$
such that $L[\rho_{\alpha\beta}] = 0$.  Hence, our algorithm finds a solution
which is a linear combination of the solutions from all the
$\mathcal{B}_{\alpha\beta}$ subspaces.  We note that physical density
matrices (with unit trace) can only exist in the diagonal sub spaces
$\mathcal{B}_{\alpha \alpha}$ due to the orthogonality between the
eigenvectors from different $\mathcal{H}_\alpha$.  However, precisely because
the unphysical density matrices from $\mathcal{B}_{\alpha\beta}$, $\alpha
\neq \beta$ have trace $0$, they can contribute to physical solutions found
by forming linear combininations with a physical density matrix, which
changes physical properties of the solution.  There are at
least $n_U$ physical, distinct NESS, which we label as $\rho^{*}_\alpha$,
where $\rho^{*}_\alpha \in B_{\alpha \alpha}$.
If another strong symmetry is present, these $n_U$ different $\rho_\alpha^*$
can be further decomposed into NESS from the new symmetry sectors.
% written as linear combinations of the corresponding density
% matrices in the various symmetry sectors of the other strong symmetry.

\paragraph{Generalization of our method for multiple NESS.---}\label{section:multiple}
We can systematically obtain all the physical steady states that
exist in all the symmetry subspaces for quantum systems with
multiple steady states, if we have knowledge of the full Lindbladian.
The simplest way would be to directly construct an ansatz that lies in the desired symmetry subspace.  If we have the capacity on the
quantum computer to generate such states, which has been
demonstrated for Dicke states~\cite{vetrivelan2022near} and states that
conserve total magnetization in the XXZ Heisenberg
chain~\cite{lyu2022symmetry}, we can simply generate such a set of states and use that to construct our hybrid ansatz for our
algorithm.  This method has the added advantage
of reducing the size of the ansatz, due to the reduction of the possible solution space.
% , as we only need to
% look for solutions in a symmetry subspace instead of the whole Hilbert
% space.  
For example, we use the quantum circuit proposed
in~\cite{lyu2022symmetry} for the eight qubit XXZ Heisenberg chain with
dephasing noise and obtained a fidelity of nearly $1$ to the theoretical
NESS in the $m = 4$ symmetry subspace with only $28$ states in our
ansatz. Here, $m$ is the eigenvalue of the total magnetization operator $M$.
However, this method is limited due to difficulty in devising circuits that conserve a general symmetry. Thus,
we also propose two general methods to find multiple NESS.

The first method utilizes the SDP structure of the optimization. For
each operator $N_k$ that corresponds to the $k$th strong symmetry in our
system, a NESS is found that is in the symmetry subspace corresponding to
a particular eigenvalue $n_k$ of $N_k$, by including the linear constraint $\text{Tr}(\beta \tilde{N_k}) = n_k$ in the SDP, where
$(\tilde{N_k})_{ij} = \bra{\chi_i} N_k \ket{\chi_j}$. These additional
linear constraints are additional, efficiently implementable, hyperplanes in the parameter space that the
optimizer needs to fulfil.

As an example, we consider
a XXZ Heisenberg chain on a system with $n$ qubits, $H_{XXZ} =
\sum_{j=1}^{n-1}\sigma_X^j\sigma_X^{j+1} + \sigma_Y^j \sigma_Y^{j+1} + \Delta \sigma_Z^j \sigma_Z^{j+1}$,
and dephasing noise, defined by the $n$ jump operators $L_i = \sigma_Z^i$. The total magnetization $M = \sum_{i=1}^n \sigma_Z^i$
commutes with the Hamiltonian and all jump operators $L_i$, generating a strong symmetry given by $S_z = e^{i\phi M}$. This gives rise to $n+1$ magnetization blocks, each associated with an eigenvalue of $M$ and has its own unique NESS.

Considering the additional constraint $\text{Tr}(\beta \tilde{M}) = m$, where $\tilde{M}_{ij} =
\bra{\chi_i} M \ket{\chi_j}$, our first method is able to obtain a solution which is in the $m$ magnetization symmetry sector of $M$ that agrees with the exact results.
% which is in agreement with the exact results.
We emphasize that the usage of the quantum computer scales linearly with the number of constraints, as we do not need to measure the $D,E,F,R$ matrices several times.

The second method does not require us to add additional constraints into
the SDP, which allows our classical post processing to be more numerically
stable. It utilizes the structure of a Vandermonde matrix to systematically remove the contributions from unwanted subspaces by applying the symmetry operator to the state and is discussed in detail in the Supplemental Information~\cite{referenceToSupplementalMaterial}.

\paragraph{Conclusion.---}
We present a new algorithm for finding NESS solutions of open systems.  Our
approach restates the NESS problem as a feasibility SDP, which is a well
known and well characterized optimization problem. We believe that this is the first work to apply this approach to
solving master equations.  As a consequence of our
approach, we are able to utilize NISQ devices to aid a classical computer in
its calculation, by offloading the difficult task of calculating expectation
values of arbitrary Pauli strings to the quantum computer.  Utilizing this
quantum assisted approach to NISQ devices, our algorithm retains all of the
advantages that such algorithms have over competitors that rely on
variationally optimizing a quantum circuit.

Our algorithm provides three main advantages over its NISQ competitors.
Firstly, since it frames the NESS problem as a feasibility SDP, it allows us
to bypass many of the problems associated with traditional variational
quantum algorithms on NISQ devices, such as the barren plateau problem and
training over the non-convex landscape in the state space.  Secondly, it
provides a natural way to enforce the positivity constraint of density
matrices during the optimization, along with any other constraints we would
want to implement.  One example where being able to enforce other constraints
is when multiple steady states exist.  Lastly, our method also gives us a
systematic way to increase the expressibility of our ansatz without
sacrificing trainability.

Our work opens up many avenues for research.  NISQ devices are already
utilized to study the ground states of chemical
substances~\cite{nam2020ground}.  Most believe that studying open system
many-body Hamiltonians, like the fermionic Hubbard model in the presence of
generic dissipations, are classically
intractable~\cite{schuch2009computational}.  It is hoped NISQ devices and
NISQ algorithms can make the simulation of such problems
possible~\cite{shtanko2021complexity}.  Our method extends these studies to
open quantum systems and widens the range of applications.  Furthermore, we
believe our method can be used as a tool to assist environmental
engineering~\cite{koch2016controlling} of open quantum systems.  Studying how
noise and ansatz choice affects quantum-assisted methods such as ours are
interesting problems to consider in the future.  We believe it is possible to
extend our algorithm to allow constraints over continuous variables, which
changes the optimization program into a semi-infinite feasibility
problem~\cite{ferrer2017comparative}.  We expect all of these to have a
substantial impact on the NESS problem in the near and far term.

\begin{acknowledgments}
\acknowledgements{We are grateful to the National Research Foundation and the Ministry of
Education, Singapore for financial support.
% The authors acknowledge the use of the IBM Quantum Experience devices for
% this work.
SV acknowledges support
from Government of India DST-SERB Early Career Research Award
(ECR/2018/000957) and Government of India DST-QUEST grant number DST/ICPS/QuST/Theme-4/2019. K.B. acknowledges funding by the DoE ASCR Accelerated Research in
Quantum Computing program (award No. DE-SC0020312), DoE QSA, NSF QLCI
(award No. OMA-2120757), NSF PFCQC program, the DoE ASCR Quantum
Testbed Pathfinder program (award No. DE-SC0019040), U.S. Department
of Energy Award No. DE-SC0019449, AFOSR, ARO MURI, AFOSR MURI, and
DARPA SAVaNT ADVENT.}
\end{acknowledgments}

% \bibliographystyle{apsrev4-1}
% \bibliography{paper}% Produces the bibliography via BibTeX.

%

\newpage
\clearpage
\onecolumngrid
\appendix

\section{Definition of cumulative $\mathcal{K}$ moment states}
\label{appendix:IQAE}
Given a Hamiltonian written as a linear combination of unitaries
\begin{gather}
    \mathcal{H} = \sum_i^r \eta_i U_i,
\end{gather}
we build an ansatz as a linear combination of quantum states
\begin{gather}
    \ket{\psi(\bm{\alpha})} = \sum_i^L \alpha_i \ket{\chi_i}.
\end{gather}
These quantum states $\ket{\chi_i}$ are chosen in a manner that utilizes the
Krylov subspace of the original problem.  With the form of the Hamiltonian
above as well as a given state $\ket{\psi}$, we build the Krylov subspace up to order $K$, defined as 
\begin{equation}
  \label{eqn: appendix krylov order k definition}
  Kr_{K}\equiv \text{span}\left\{
  \ket{\psi},H\ket{\psi},\cdots,H^{K}\ket{\psi}\right\}
\end{equation}
We take each element of the subspace
$H^k\ket{\psi}=(\sum_i\eta_i U_i)^k\ket{\psi}$ and multiply out the power $k$
such that we get a sum. 
Each constituent term of this sum, which can be written as $U_{i_{1}}\dots U_{i_k}\ket{\psi}$, is then added to
a set $\mathbb{S}_k$.  Finally, we combine all $K+1$ sets $\mathbb{S}_k$ into
what we now call the fine-grained Krylov subspace basis or cumulative
$K$-moment states $\mathbb{CS}_{K}\equiv\cup_{j=0}^{K}\mathbb{S}_{j}$, which
is formally defined as follows~\cite{bharti2021iterative}.
\begin{defn}  \label{def:cumulant_states}
Given a set of unitaries $\mathbb{U}\equiv\left\{ U_{i}\right\}_{i=1}^{r}$, a
positive integer $K $ and some quantum state $\vert\psi\rangle,$ $K$-moment
states is the set of quantum states of the form $ \{\ket{\chi}\}_K =\left\{
U_{i_{K}}\cdots U_{i_{2}}U_{i_{1}}\vert\psi\rangle\right\} _{i}$ for
$U_{i_{l}}\in\mathbb{U}.$ We denote the aforementioned set by
$\mathbb{S}_{K}$.  The fine-grained Krylov subspace basis or cumulative
$K$-moment states $\mathbb{CS}_{K}$ is defined as
$\mathbb{CS}_{K}\equiv\cup_{j=0}^{K}\mathbb{S}_{j}$.
\end{defn}

The success of such an ansatz for the NESS problem is highly dependent on the choice of the initial state $\ket{\psi}$ and also on the steady state being low rank. For the use case discussed here, it is generally hoped that the initial state has decent overlap with the steady state. If this is done, we expect that a sufficient ansatz can be generated for low $k$. It is known that the Hamiltonian ground state problem is QMA-hard
%problem of finding an exact quantum ground state for an arbitrary Hamiltonian is QMA-hard, 
and finding the steady state of a Liouvillian is probably of similar complexity. For most prior algorithms developed for NISQ computers for finding the ground state, they rely very heavily on utilizing ansatz that have some guarantee of overlap with the true ground state, of which without, it would not be feasible. Similarly, if we start out with a initial state that has exponentially small overlap with the true steady state, this method of generating an ansatz will not be efficient.

A variant of the $\mathbb{C}\mathbb{S}_K$ ansatz that we use in the main text for the five and eight qubit transverse field Ising model is to define $\mathbb{CS}_{K}\equiv\cup_{j=0}^{K}\mathbb{S}^\prime_{j}$
where $\mathbb{S}^\prime_1$ is a random subset with cardinality $q$ of $\mathbb{S}_1$, and $\mathbb{S}^\prime_j$ is defined as a random subset with cardinality $q$ of the set $\{U_i \ket{\phi} | U_i \in \mathbb{U} \text{ and } \ket{\phi} \in \mathbb{S}_{j-1}\}$. This variant of the $\mathbb{C}\mathbb{S}_K$ ansatz has a smaller cardinality than the $\mathbb{C}\mathbb{S}_K$ ansatz defined above, at a cost of lower expressibility. For this particular variant of the $\mathbb{C}\mathbb{S}_K$ ansatz, it is more meaningful to quantity the expressibility of $\mathbb{C}\mathbb{S}_K$ by using the number of ansatz states in each $\mathbb{C}\mathbb{S}_K$ rather than using the index $K$ itself.

\section{Description of feasibility SDP}\label{appendix:feasibility_SDP}
The typical Semidefinite programming (SDP) problem is one that minimizes a linear function of a cost function subject to the positive semi-definite condition and additional $m$ constraints

\begin{gather}
    \min_\beta \text{Tr} (A \beta),\\
    \text{subject to } \beta \succcurlyeq 0,\\ \text{Tr}(C_i \beta) = c_i, i = 1,\dots, m.
\end{gather}

We note that although the standard form for these additional $m$ constraints is to be expressed as a trace constraint, this is not necessary. The works of Boyd, along with~\cite{wolkowicz2012handbook}, offer alternate forms of SDPs that are not formulated with trace constraints.

In this paper, instead of formulating the problem as a typical SDP, it is formulated as a feasibility SDP problem~\cite{boyd2004convex}. In this problem, the cost function is removed. The program now attempts to find if it is possible to find a $\beta$ subject to the positive semi-definite condition and the additional $m$ trace constraints. Once the algorithm finds a $\beta$, the program stops and reports the solution. There is hence no concept of a cost function to be minimized. The exact complexity of this problem is still unknown, though it is highly suspected that the feasibility SDP is polynomial in the size of the matrix~\cite{ramana1997exact}. Furthermore, it has been shown that certain classes of feasibility SDPs still can be efficiently solved~\cite{kalantari2019equivalence}. 

The feasibility SDP problem can be solved with existing SDP solvers, by setting the cost function to $0$. As mentioned in the main manuscript, we rely on CVX to solve the SDPs in this work.

\section{Summary of the algorithm}
\label{appendix:algorithm_steps}
We can summarise our algorithm as a three step procedure, namely
\begin{enumerate}
  \item{Write the ansatz $\rho=\sum_{i,j}\beta_{ij}\ket{\chi_i}\bra{\chi_j}$,
  where $\ket{\chi_i}\in\mathbb{C}\mathbb{S}_K$.  With this ansatz, the
  Linblad master equation is given by
  \begin{gather}
    -i(D\beta E-E\beta D)
      + \sum_n \gamma_n \left(R_n \beta R_n^\dag
      -\frac{1}{2}F_n \beta E -
      \frac{1}{2} E \beta F_n\right) = 0.
  \end{gather}
  where the matrices $D,E,R_n,F_n$ have matrix elements $E_{ij} =
  \braket{\chi_i|\chi_j}$, $D_{ij} = \mel{\chi_i}{H}{\chi_j}$, $R_{(n),ij} =
  \mel{\chi_i}{A_n}{\chi_j}$ and $F_{(n),ij} = \mel{\chi_i}{A_n}{\chi_j}$}
  \item{Measure the matrix elements of the $D,E,R_n,F_n$ matrices on a
  quantum computer.}
  \item{Solve the following feasibility SDP on a classical computer, written as 
  \begin{subequations}
  \label{eqn:feasibility_sdp_appendix}
  \begin{gather}
      \text{Find } \beta \text{ s.t.}
      -i(D\beta E- E \beta D)
      + \sum_n \gamma_n \left(R_n \beta R_n^\dag
      -\frac{1}{2} F_n \beta E -
      \frac{1}{2} E \beta F_n\right) = 0,\\
      \beta \succcurlyeq 0,\\
      \text{Tr}(\beta E) = 1.
  \end{gather}
\end{subequations}}
\end{enumerate}

\section{Extra results for 5 qubits}
\label{appendix:extra5qubit}

\begin{figure}[H]
    \centering
    \includegraphics[width=0.49\textwidth]{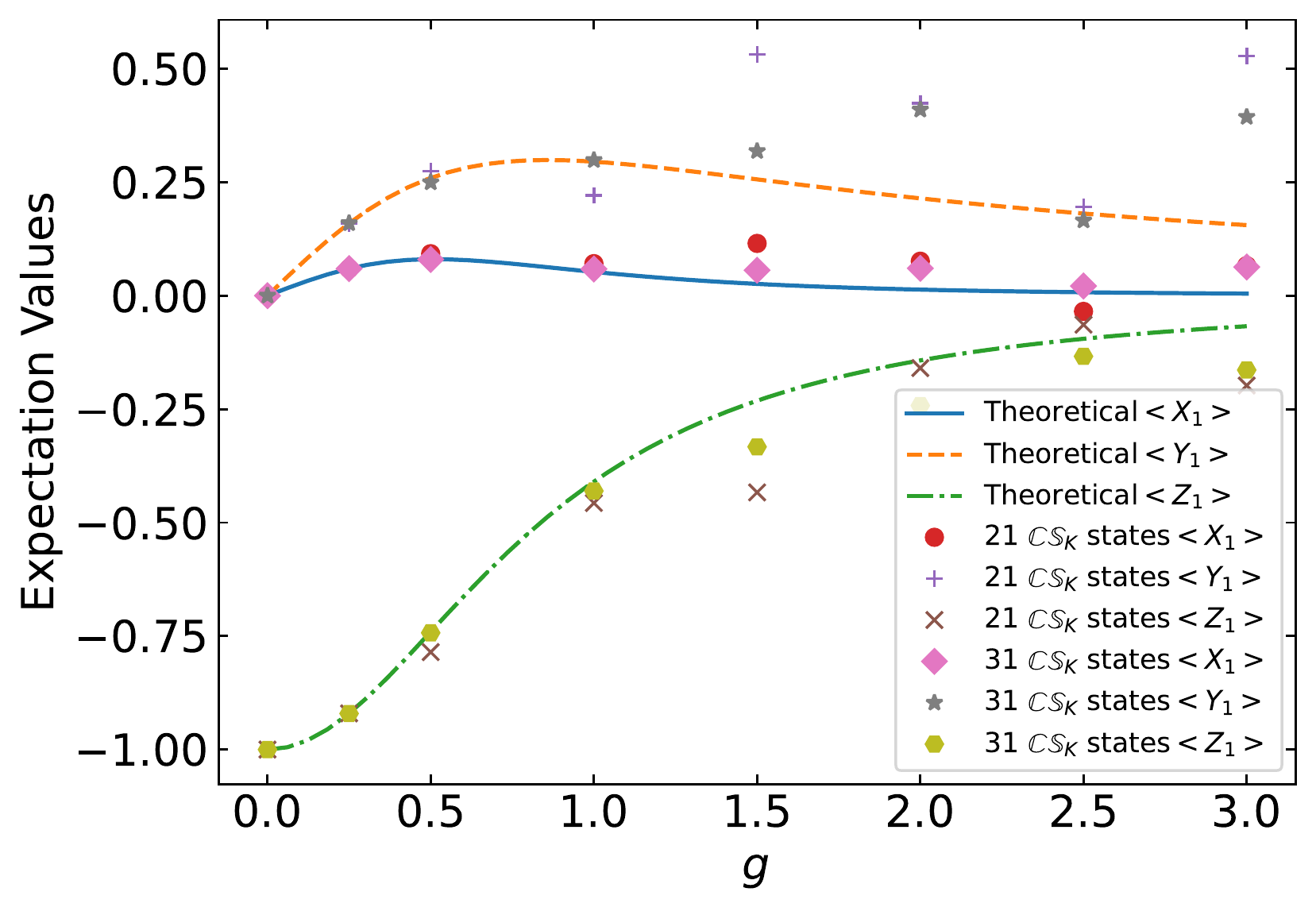}
    \caption{Presented here are extra results for the 5-qubit model to supplement the results in the main manuscript. Shown are the expectation values of various
    observables for the steady states we obtained for the five qubit transverse field Ising model ($\gamma=1$).}
    \label{fig:5_qubit}
\end{figure}

\section{Overlap of initial state used to generate $\mathbb{C}\mathbb{S}_K$ states for 8 qubits}
\label{appendix:8qubitinitialstate}

The effectiveness of the $\mathbb{C}\mathbb{S}_K$ method to generate basis states for the ansatz relies heavily on choosing an initial state with a non-exponentially decreasing overlap with the exact NESS. In general, we believe this is hard, due to the fact that it is probably closely related to the ground state problem, which is QMA-hard. In our simulations for $8$ qubits, we rely on using a starting state that is an eigenstate with the largest eigenvalue of the exact NESS. In Table \ref{tableoverlap} we show the overlap of such a state with the exact NESS. Even with such a method, for large $g$, this is not very effective in producing a good steady state. We attribute the need for larger ansatz sizes in the large $g$ case to this reason.

\begin{table*}[h]
\centering
\begin{tabular}{|l|l|l|l|l|l|l|l|l|}
\hline
              & $g=0$ & $g=0.25$ & $g=0.5$ & $g=1$ & $g=1.5$ & $g=2$ & $g=2.5$ & $g=3$ \\ \hline
Overlap  & 1     &  0.811     &  0.469     & 0.123      &  0.0474     &    0.0263   & 0.0181   & 0.0141  \\ \hline

\end{tabular}
\caption{Comparison of the fidelity of the starting states with the exact NESS solution that were used to construct the $\mathbb{C}\mathbb{S}_K$ ansatz for the 8 qubit example discussed in the main manuscript. As $g$ increases, we see that it is harder to find good starting states with non-vanishing overlap with the actual NESS.}
\label{tableoverlap}
\end{table*}

\section{Extra results for 8 qubits and detailed comparison with existing approach}
\label{appendix:extra8qubit}
\begin{figure}[H]
    \centering
    \includegraphics[width=0.80\textwidth]{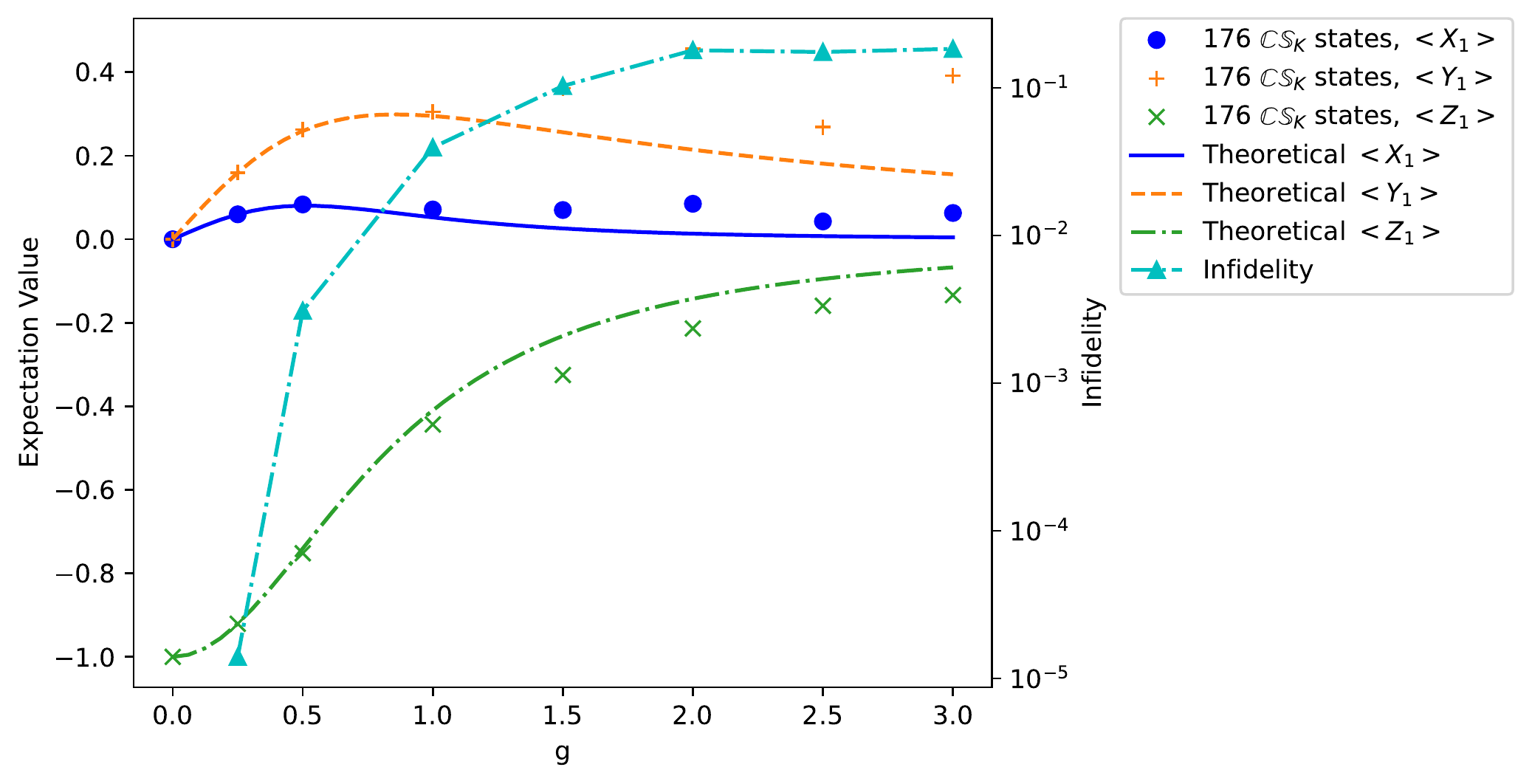}
    \caption{Presented here are extra results for the 8-qubit model to supplement the results in the main manuscript. Shown are the expectation values of various
    observables for the steady states we obtained for the eight qubit transverse field Ising model ($\gamma=1$). The ansatz size used in this simulation was a $\mathbb{C}\mathbb{S}_K$ ansatz comprising of $176$ constituent states. }
    \label{fig:8_qubit}
\end{figure}

Here, we present extra results for the $8$ qubit transverse field Ising
model for easy comparison to the currently existing Variational Quantum
Algorithm (VQA)-based approach in~\cite{yoshioka2020variational}.  This is
shown in \figref{fig:8_qubit}.  We also perform a detailed resource
comparison of both our approach and the VQA-based approach
in~\cite{yoshioka2020variational} based on what was mentioned in
Appendix~\ref{appendix:comparison}.
First, we list down the resource requirements of the VQA-based approach
in~\cite{yoshioka2020variational}.  Each evaluation of the cost function
$\langle \rho(\vec{\theta}) | \mathcal{L}^\dagger \mathcal{L}|
\rho(\vec{\theta}) \rangle$ in their VQA requires the evaluation of $1713$
$16$-qubit quantum circuits.  Assuming that their classical-quantum feedlack
loop requires $N$ evaluations of the cost function, this means that they need
to evaluate at least $1713 \times N$ $16$-qubit quantum circuits.  $N$ is in
general a large number, since the classical VQA optimisation is shown to be
NP-hard~\cite{bittel2021training} and also due to the well-known issue with
barren plateaus that VQAs face.
% In our own numerical experiments using the classical BFGS optimiser, we did
% not achieve convergence even after $N = 900$, which means that from our own
% numerical experiments, at the very least, we need to evaluate $1541700$
% $16$-qubit circuits.

On the other hand, for our SDP based method, we need to evaluate the matrix
elements $E_{ij}$, $D_{ij}$, and $R_{ij}$, $F_{ij}$ for both the
phase-damping and ampltitude damping dissipators.  Assuming that we have $M$
states in our ansatz, this means that in the worst case we need to measure
$20\times(M^2 + M) - M$ $8$-qubit quantum circuits.  Usually, we need less
measurements than that because there would be matrix elements that are
repeated, based on the ansatz that we chose.  Recall that once we have
measured the matrix elements on the quantum computer, we don't need the
quantum computer anymore.  What remains is the classical SDP optimisation
which is convex and hence classically tractable.  As can be seen in
\figref{fig:8_qubit}, for small values of $g$, we can get good fidelity with
$M = 176$, which means that we need to evaluate about $622864$ $8$-qubit
quantum circuits.

Based on the above analysis, as long as the VQA method requires more than $N
\approx 360$ evaluations of the cost function, our SDP method (in the worst
case) outperforms the VQA method in terms of number of measurements required.
Furthermore, to reiterate, our method requires the measurement of $8$-qubit
quantum circuits, whereas the VQA method in \cite{yoshioka2020variational}
requires the measurement of $16$-qubit quantum circuits.  We can give a lower
bound on the number of cost function evaluations in the VQA approach with the
following argument.  In the minimization of the cost function with gradient
descent with the VQA approach, asusming a fixed step size, we need
$O(\frac{1}{\epsilon})$ number of steps at least for the cost function to be
accurate up to $\epsilon$.  Hence, assuming we set $\epsilon = 0.01$, we need
at least 100 steps in the gradient descent.  With $47$ parameters in the
quantum circuit in \cite{yoshioka2020variational}, each step of the gradient
descent algorithm requires at least $47$ evaluations of the cost function,
since the computation of the gradient at each step requires us to
independently vary each parameter at least once.  Thus, at least $47$
evaluations of the cost function at each step for at least $100$ steps means
that the VQA approach requires at least $N = 4700$ evaluations of the cost
function to achieve an accuracy up to $\epsilon = 0.01$.  This lower bound
already exceeds $N = 360$, which means that the lower bound of the amount of
quantum resources required to get the cost function to be accurate up to
$0.01$ already exceeds the amount of quantum resources required on our end.
We note that the argument to derive this lower bound does not take into
account the NP-hardness of the VQA optimisation and the issue with Barren
Plateaus, both of which would increase $N$ significantly.  In our own
numerical experiments of the VQA method using the classical BFGS optimiser,
we did not achieve convergence even after $N = 900$.  This means that when we
compare both methods, our method requires a lot lesser measurements of
$8$-qubit quantum circuits in the worst case than the amount of measurements
on $16$-qubit quantum circuits required by~\cite{yoshioka2020variational} in
the best case.

% we need a lot lesser measurements of $8$-qubit quantum
% circuits than 
% In our own numerical 
% All this shows that as long as $N > 363$, which we find to be the case in our own numerical experiments using the classical BFGS optimiser

\section{Comparison to alternative approaches}
\label{appendix:comparison}

VQAs have been investigated in detail, and some of the drawbacks of using
VQAs include the existence of barren
plateaus~\cite{mcclean2018barren,grant2019initialization}, the classical
quantum feedback loop being a major bottleneck on current cloud-based quantum
computers, and finally the fact that the classical optimisation program
does not belong to any class of mathematically well-studied programs.
Furthermore, recent work has shown that VQAs a large number of parameters to
be useful \cite{larocca2021theory, anschuetz2021critical}.  Some VQAs also
require complicated multi-qubit controlled unitaries.  Hence, the application
of VQAs to study NESS in \cite{yoshioka2020variational} inherits all of these
aforementioned drawbacks.  Furthermore, the work in
\cite{yoshioka2020variational} requires $n$ ancilla qubits to solve an $n$
qubit problem due to their approach of mapping an $n$ qubit density matrix to
a $2n$ qubit statevector.  Lastly, it is unclear how the positivity of the
density matrix is maintained over the course of the variational optimisation.
This problem is possibly a difficult one, since it is known that checking the
positivity of a density matrix in tensor network methods is
NP-hard~\cite{kliesch2014matrix}.

The canonical example of a VQA is the variational quantum eigensolver
(VQE)~\cite{peruzzo2014variational,kandala2017hardware}, which aims to solve
for the ground state of a given Hamiltonian.  Other examples of areas where
VQAs have found application include quantum
simulation~\cite{li2017efficient,yuan2019theory,benedetti2021hardware},
finding of excited states~\cite{higgott2019variational}, quantum
thermalization~\cite{verdon2019quantum}, numerical
solvers~\cite{anschuetz2019variational,bravo2020quantum,huang2019near}, to
name a few.

As shown in our Table \ref{table:comparison}, our algorithm addresses some of
the challenges in VQA/VQE based methods such as
in~\cite{yoshioka2020variational}.

We would also like to mention that for VQA based methods, the problem of how
to systematically obtain an ansatz for such algorithms that is expressible
enough to approximate the solution to a desired accuracy, yet retain
trainability, is inherent and might be insurmountable.  It has been shown
that to get an expressible enough ansatz in such a framework, demands a
fundamental tradeoff with the trainability of such an
ansatz~\cite{holmes2021connecting}.

\begin{table*}[]
\centering
\caption{Comparison of our solution to NESS problem to other solutions.}
\resizebox{0.99\textwidth}{!}{%
\begin{tabular}{|p{0.11\linewidth}|p{0.35\linewidth}|p{0.35\linewidth}|}
\hline
                       & VQE based methods~\cite{yoshioka2020variational}                                                                       & Our algorithm                                                                                                                                                                                           \\ \hline
Ansatz &    Parameterized quantum state ansatz. &       Hybrid density matrix ansatz.              
                       
                       \\ \hline
Variational parameters & Utilizes a parameterized quantum state/circuit. Variational parameters modify the quantum state on the quantum computer. & Utilizes a classical combination of quantum states. Variational parameters are the coefficients defining the linear superposition of the quantum states of the ansatz, do not modify the quantum state on the quantum computer. \\ \hline
Feedback loop          & Requires a classical-quantum feedback loop, constantly modifying quantum state on the quantum computer.                            & Once measurements are made on quantum computer, no longer require usage of quantum computer, no need for classical-quantum feedback loop.                                                                                      \\ \hline
Training landscape     & Non-convex landscape, optimization shown to be NP-hard, have exponentially many local minima far away from global minima.                       & Convex landscape.                                                                                                                                                                                                              \\ \hline
Ansatz selection       & Usually rely on problem agnostic ansatz such as Hardware Efficient ansatz, no guarantee of systematic improvement.                 & Problem aware ansatz. Systematic way to improve without sacrificing trainability.                                                                                                                                                                               \\ \hline

Optimization program & Generally does not belong in any mathematically
well studied class of programs. & Feasibility SDP, mathematically well
characterized and studied.

\\ \hline
Positivity constraint  & Not aware of how to systematically enforce positivity constraint for density matrix.                                               & Systematic way to enforce, built into optimization program.                                                                                                                                                                    \\ \hline
Multiple NESS solutions & Not clear how to find multiple solutions in the
presence of strong symmetries. & Systematic ways to find multiple solutions
exist, such as adding linear constraints into the SDP optimisation.  \\
\hline

\end{tabular}%
}
\label{table:comparison}
\end{table*}

\section{Justification for ansatz, and scaling arguments}
\label{appendix:scaling}

The scaling of our algorithm is fundamentally related to the 
problem of obtaining an appropriate ansatz that is expressible
enough.  It is known that to prepare an arbitrary state on an $n$ qubit
quantum computer, we require a circuit depth of at least
$2^n/n$~\cite{knill1995approximation,mottonen2004quantum,vartiainen2004efficient,plesch2011quantum}.
This is a complexity theoretic statement that cannot be bypassed by any
quantum simulation algorithm based on parametric quantum circuits or linear
combination of quantum states, and indicates that our algorithm shares the worst case of requiring an exponentially large ansatz to obtain perfect fidelity with other variational algorithms, in the case of a general Hamiltonian with no symmetries and which is ergodic.

Thus, a large contributing factor to how our algorithm will scale hinges on the
manner which we choose the $\ket{\chi_i}$ states which we generate our hybrid
density matrix ansatz with.  As such, a justification for the manner which
we generate those states will be provided in this appendix (first given in
\cite{bharti2021iterative}), before we discuss the scaling of our algorithm.
Our justification relies on similar ideas to the method of using imaginary
time evolution to find the ground state on classical computers.  Suppose the
initial state ($0$-moment state) $\vert\psi\rangle$ can be expressed in the
eigenbasis of the Hamiltonian $H$,

\begin{equation}
\vert\psi\rangle=\sum_{i=1}^{\mathcal{N}}a_{i}\vert\phi_{i}\rangle\,
,\label{eq:initial_state_eigen_decomposition_ap}
\end{equation}
where $a_{i}\in\mathbb{C}$ for $i\in\{1,2,\cdots,\mathcal{N}\}$.  Now we
consider the normalized state if we apply the operator $\exp\left(-\tau
H\right)$
for some $\tau\geq0$ on the initial state:
\begin{equation}
\vert\gamma\rangle=\frac{e^{-\tau H}\vert\psi\rangle}{\sqrt{\langle\psi\vert
e^{-2\tau H}\vert\psi\rangle}}.\label{eq:ITE_state_1_ap}
\end{equation}
In normal imaginary time evolution, this is the point where we recognize that
if $\tau\rightarrow\infty$, $\ket{\gamma}\rightarrow\ket{E_0}$, where
$\ket{E_0}$ is the ground state.  We consider the power series expansion
$e^{-\tau H}=\sum_{p=0}^{\infty}\frac{\left(-\tau H\right)^{p}}{p!}$,
\begin{equation}
\vert\gamma\rangle=\frac{\sum_{p=0}^{\infty}\frac{\left(-\tau
H\right)^{p}}{p!}\vert\psi\rangle}{\sqrt{\langle\psi\vert\sum_{p=0}^{\infty}\frac{\left(-2\tau
H\right)^{p}}{p!}\vert\psi\rangle}}.\label{eq:ITE_state_expanded_ap}
\end{equation}
Let us also define the operator 
\begin{equation}
\mathcal{O}^{K}\equiv\sum_{p=0}^{K}\frac{\left(-\tau H\right)^{p}}{p!},\label{eq:K-Approx_Operator-1_ap}
\end{equation}
for $K\geq0.$ We note that $\mathcal{O}^{K}$ corresponds to the sum of first
$K$ terms of $e^{-\tau H}$, or in other words a sum of elements of the Krylov
subspace of $H$ and $\ket{\psi}$ up to $K$, where the Krylov subspace is
defined as per \eqref{eqn: appendix krylov order k definition}.
Using $\mathcal{O}^{K},$ we
proceed to define
\begin{equation}
  \vert\gamma_{K}\rangle\equiv\frac{\sum_{p=0}^{K}\frac{\left(-\tau
  H\right)^{p}}{p!}\vert\psi\rangle}{\sqrt{\langle\psi\vert
  \left(\sum_{p=0}^{K}\frac{\left(-\tau
  H\right)^{p}}{p!}\right)^2\vert\psi\rangle}}.\label{eq:Evolved_state_ITE_K-Approx-1_ap}
\end{equation}
For $K\rightarrow\infty$, $\vert\gamma_{K}\rangle\rightarrow\vert\gamma\rangle.$
Remembering that we are expressing the Hamiltonian as a linear combination of
unitaries, it is easy to see that $\vert\gamma_{K}\rangle$ can be written as
linear combination of cumulative $K$-moment states, i.e,
\[
\vert\gamma_{K}\rangle=\sum_{\vert
\chi_{i}\rangle\in\mathbb{CS}_{K}}\alpha_{i}\vert \chi_{i}\rangle
\]
where the combination coefficients $\alpha_{i}\in\mathbb{C}.$ The
aforementioned arguments justify our choice of ansatz as a linear
combination of cumulative $K$-moment states.

In the worst case, the number of overlaps scales as $O(r^K)$ for $r$ terms in
$H$.  This is fundamentally an expressibility problem, present in all NISQ
variational algorithms, be it based on linear combination of states or those
based on parametric quantum circuits.  It is known that to prepare an
arbitrary state on an $n$ qubit quantum computer, we require a circuit depth
of at least
$2^n/n$~\cite{knill1995approximation,mottonen2004quantum,vartiainen2004efficient,plesch2011quantum}.
This suggests that it is very hard to produce an expressible enough ansatz to
reproduce an arbitrary quantum state in the Hilbert space.

\section{An algorithm to find multiple steady states}
\label{appendix:multiple_NESS}

Before we proceed, we make the assumption that our symmetry operator $U$ can
be written as a linear combination of $r$ tensored-Pauli operators $P_i$, namely
\begin{equation}
  \label{eqn: symmetry operator LCU}
  U = \sum_{i=0}^r \gamma_i P_i, \quad \gamma_i \in \mathbb{C}.
\end{equation}
Under this assumption, it is easily to calculate powers of $U$, i.e $U^k$ can be expanded in the same Pauli basis as
$U^k = \sum_{i=0}^{r^\prime} \gamma_i^\prime P_i$.
Hence, given a $\rho$ of the form used in the paper, we have
\begin{align}
  \tr{\rho U^k} 
  &= \sum_{m=0}^{r^\prime} \gamma_m^\prime \sum_{ij}\beta_{ij}
  \mel{\chi_i}{P_m}{\chi_j} \nonumber \\
  \label{eqn: trace rho k}
  &= \sum_{m=0}^{r^\prime} \gamma_m^\prime \tr{\beta Q_m}.
\end{align}
where $Q_m$ is the matrix with matrix elements $\mel{\chi_i}{P_m}{\chi_j}$.
Now, we can efficiently calculate the matrix elements
$\mel{\chi_i}{P_m}{\chi_j}$ on the quantum computer since $\ket{\chi_i}\in
\mathbb{C}\mathbb{S}_K$~\cite{bharti2020quantum,bharti2021iterative}.  Hence,
we see that once we have the $\beta$ that corresponds to $\rho$, we can
easily compute terms like $\tr{\rho U^k}$ on the quantum computer by
measuring $\mel{\chi_i}{P_m}{\chi_j}$ on the quantum computer.
The above argument can be similarly extended to show that for an
observable $O$ written as a linear combination of tensored-Pauli operators,
we can easily compute terms like $\tr{\rho O U^k}$, $\tr{ U^k \rho O}$, $\tr{
U^k \rho U^{k^\prime} O}$ on the quantum computer once we have $\beta_{ij}$.

We illustrate the main idea here with a simple, pedagogical example.
Assume there exists a symmetry operator $U$ with eigenvalues $\pm 1$.  This gives rise to 4 symmetry sectors in $\mathcal{B}$,
$\mathcal{B}_{++},\mathcal{B}_{+-},\mathcal{B}_{-+},\mathcal{B}_{--}$ with
corresponding density matrices $\rho_{++},\rho_{+-},\rho_{-+},\rho_{--}$ such
that $L[\rho_{\alpha,\beta}] = 0, \alpha,\beta = {+,-}$.  Here, $\rho_{++}$
and $\rho_{--}$ correspond to physical density matrices and are the two
NESS, while $\rho_{+-}$ and $\rho_{-+}$ are traceless and unphysical.
We first obtain a single solution $\rho$
by solving the SDP.  This solution is a linear combination, $\rho = a\rho_{++}+b\rho_{+-}+c\rho_{-+}+d\rho_{--}$.
We note that $\rho^{\prime} = U\rho U^\dagger =
a\rho_{++}-b\rho_{+-}-c\rho_{-+}+d\rho_{--}$, which implies $\rho^{\prime
\prime} = \frac{1}{2}\left(\rho + \rho^\prime\right) =
a\rho_{++}+d\rho_{--}$.  We thus obtain $\rho_{++}$ ($\rho_{--}$) by
evaluating $(\rho^{\prime \prime}+(-)U \rho^{\prime \prime})/2$.  
% Lastly, as
% long as $U$ can be written as a linear combination of unitaries, we can
% compute expectation values using the $\rho^{*}_\alpha$ that we obtain with
% this procedure.

% Here, we consider an example where there are two strong symmetries in
% our system to show that our approach is amenable to the case where there are
% multiple strong symmetries in our system.  For this example, we will use the
% second method in conjuction with the first method described above.
As an example, we drive the aforementioned $n$ qubit $H_{XXZ}$ with two non-local Lindblad jump operators, namely
$A_{XXZ}^1 = \sqrt{\Gamma(1-\mu)}\sigma_+^1\sigma_-^{n}$ and $A_{XXZ}^2 =
\sqrt{\Gamma(1+\mu)}\sigma_-^1\sigma_+^{n}$,
where $\sigma_\pm^j = (\sigma_X^j \pm i \sigma_Y^j)/2 $, $\Gamma > 0$ and $\mu \in [0,1]$~\cite{manzano2018harnessing}. There are two unitary symmetry operators, $S_z=e^{i\phi M}$  and $S \equiv P \prod_{j=1}^{n} \sigma_X^j$,
where $P$ is the operator that exchanges site $j$ with site $L-j+1$ for all
$j$.  This can be defined using the computational basis.  If the eigenvectors
of $\sigma_Z^j$ are expressed as $\ket{s_1,\dots,s_n}$, where $s_j \in \{0,1\}$, $P$
is defined as $P \equiv \sum_{(s_1,\dots,s_n)\in \{0,1\}^n}\ket{s_1,\dots,s_n}\bra{s_n,\dots,s_1}$. We can show that $[S,H_{XXZ}]=[S,A_{XXZ}^1]=[S,A_{XXZ}^2]=0$, and also that
$S$ has two eigenvalues $(\pm 1)$.  Thus we would expect four different invariant subspaces, of which only two hold physical steady states, and the other two subspaces can contribute as components of a physical solution. We start with a solution obtained from method one that is in a selected symmetry subsection of $M$, that has contributions from all the steady state solutions of $S$. The second method can next be applied to remove the
contributions from unwanted symmetry sectors of $S$.Our approach was tested
up to system sizes of eight qubits, to find two trace orthogonal solutions
($\text{Tr}(\rho_1^\dagger\rho_2)=0$) in the zero magnetization symmetry
sector of $M$, which each solution corresponding to a physical steady state
of the symmetry operator $S$.  Our algorithm gave results that were in
agreement with the exact results.

Now to describe the method in more detail, denoting $\rho_\alpha^*$ as the physical NESS belonging to a particular
symmetry subspace, the goal is then to obtain all of the $\rho_\alpha^*$, which is equivalent to finding all of the corresponding
$\beta_\alpha^*$, since once we do so, we can compute expectation values of
the form $\tr{O\rho_\alpha^*}$.  For simplicity, we shall assume below that
there is only $1$ strong symmetry (only one symmetry operator $U$), though
the method detailed below easily generalises for multiple symmetries.
Without loss of generality, let the operator $U$ have $n_U$ distinct
eigenvalues which we know and label as $e^{i\lambda_1}, \dots,
e^{i\lambda_{n_U}}$.  As mentioned in the main text,
%  section~\ref{section:multiple},
running our algorithm once gives us a $\rho^{(1)}$ that is the following
linear combination:
\begin{equation}
  \label{eqn: initial linear combination}
  \rho^{(1)} = \sum_{\alpha,\beta}c_{\alpha,\beta}^{(1)}\rho_{\alpha,\beta},
  \quad \rho_{\alpha,\beta}\in \mathcal{B}_{\alpha,\beta}
\end{equation}
where the coefficients $c_{\alpha,\beta}^{(1)}$ are unknown to us.  Here,
$\rho_{\alpha,\beta}$ are the density matrices such that
$L[\rho_{\alpha,\beta}] = 0$. The general method is to first
systematically eliminate the contributions from the subspaces
$\mathcal{B}_{\alpha\beta}$, $\alpha \neq \beta$ in the above linear
combination. To do so, we note that
\begin{equation}
  U\rho^{(1)}U^\dagger =
  \sum_{\alpha}c_{\alpha,\alpha}^{(1)}\rho_{\alpha,\alpha} +
  \sum_{\alpha\neq\beta} e^{i(\lambda_\alpha - \lambda_\beta)}
  c_{\alpha,\beta}^{(1)}\rho_{\alpha}{\beta}
\end{equation}
which means that  
\begin{equation}
  \rho^{(1)} - U\rho^{(1)}U^\dagger = \sum_{\alpha \neq
  \beta}\left(1-e^{i(\lambda_\alpha - \lambda_\beta)}\right)
  c_{\alpha,\beta}^{(1)}\rho_{\alpha}{\beta}
\end{equation}
Now, to eliminate the contribution from the $\mathcal{B}_{m,n}$, $m \neq n$
subspace, we define $\rho^{(2)}$ as follows:
\begin{equation}
  \label{eqn: second step linear combination}
  \rho^{(2)} = \rho^{(1)} - \frac{(\rho^{(1)} -
  U\rho^{(1)}U^\dagger)}{1-e^{i(\lambda_m - \lambda_n)}} =
  \sum_{\substack{\alpha,\beta \\\alpha \neq m\\ \beta \neq n}}
  c^{(2)}_{\alpha,\beta} \rho_{\alpha,\beta}
\end{equation}
where 
\begin{equation}
  c^{(2)}_{\alpha,\beta} = 
  \begin{cases}
    c^{(1)}_{\alpha,\alpha} \quad \text{if }\alpha = \beta \\
    \left(1-\frac{1-e^{i(\lambda_\alpha -
    \lambda_\beta)}}{1-e^{i(\lambda_m -
    \lambda_n)}}\right)c^{(1)}_{\alpha,\beta} \quad \text{if }\alpha \neq \beta
  \end{cases}
\end{equation}
We see that even though we do not know the coefficients in the linear
combinations in \eqref{eqn: initial linear combination}, we can
systematically eliminate the contribution from $\mathcal{B}_{m,n}$ by
considering linear combinations of the terms $\rho^{(1)}$ and $U
\rho^{(1)}U^\dagger$.  Now, since \eqref{eqn: second step linear
combination} is of the same form as \eqref{eqn: initial linear combination},
but just without the contribution from the subspace $\mathcal{B}_{m,n}$, we
can pick another subspace $\mathcal{B}_{o,p}$ with $o\neq p$ and follow the same
process to get a density matrix $\rho^{(3)}$ but with the contribution from
$\mathcal{B}_{o,p}$ eliminated.  It is easy to see that we can continue this
process until we have a density matrix $\rho^{\text{phys}}$ which only
contains linear combinations of physical density matrices, i.e 
\begin{equation}
  \label{eqn: rho phys}
  \rho^{\text{phys}} = \sum_{\alpha=1}^{n_U} c_\alpha \rho_{\alpha,\alpha} 
\end{equation}
Note that at this stage, we have $\rho^{\text{phys}}$ written as a linear
combination of terms $\rho^{(1)}$, $U\rho^{(1)}U^\dagger$,
$U^2\rho^{(1)}(U^\dagger)^2$, etc, where $\rho^{(1)}$ is the solution that
our algorithm originally gave us.
Now, with \eqref{eqn: rho phys}, we can obtain $c_\alpha
\rho_{\alpha,\alpha}$ as a linear combination of the matrices
$\rho^{\text{phys}}$, $U\rho^{\text{phys}}$, \dots
$U^{n_U-1}\rho^{\text{phys}}$. To see this, we note that we have 
\begin{equation}
  \begin{pmatrix}U^0\rho^{\text{phys}} \\ U^1\rho^{\text{phys}} \\ \vdots\\
  U^{n_U-1}\rho^{\text{phys}} \end{pmatrix} = V \begin{pmatrix}c_1 \rho_{11}
  \\ c_2 \rho_{22} \\ \vdots\\
  c_{u_N} \rho_{22} \end{pmatrix}
\end{equation}
where 
\begin{equation}
  V = \begin{pmatrix} 1& 1& \dots& 1 \\
   e^{i\lambda_1} &e^{i\lambda_2}&\dots &e^{i\lambda_{n_U}} \\
  \vdots&\vdots&\vdots&\vdots\\ 
  e^{i(n_U-1)\lambda_1}&e^{i(n_U-1)\lambda_2}& \dots &e^{i(n_U-1)\lambda_n}
 \end{pmatrix}
\end{equation}
is the Vandermonde matrix~\cite{robinson2006course}.  Since
$e^{i\lambda_\alpha}$ are the $n_U$ distinct eigenvalues of $U$, the
Vandermonde matrix is invertible.  Once we have $c_\alpha
\rho_{\alpha,\alpha}$, we can just normalise it by its trace to give us a
density matrix $\rho_{\alpha,\alpha} \in \mathcal{B}_{\alpha,\alpha}$ such
that $L[\rho_{\alpha,\alpha}] = 0$.  Note that at this stage, we have
$\rho_{\alpha,\alpha}$ as a linear combination of terms like $U^{k}\rho^{(1)}
U^{k^\prime}$ for some $k,k^\prime \in \mathbb{Z}^+$, where $\rho^{(1)}$ is the
solution that our SDP algorithm gave.  For an observable $O$ written as a
linear combination of tensored-Pauli operators, we can easily obtain
$\tr{\rho_{\alpha,\alpha}O}$ in a similar fashion as in \eqref{eqn: trace rho
k} by using the fact that $O$ and $U$ are both linear combination of
tensored-Pauli operators, and hence in the end, we only need to evaluate
terms like $\mel{\chi_i}{P}{\chi_j}$ on the quantum computer, where $P$ is a
tensored-Pauli operator.  We note that the method described above to get
$\rho^{*}_\alpha$ works only if the coefficient
$c^{(1)}_{\alpha,\alpha}$ in $\rho^{(1)}$ is not zero.  Since $\rho^{(1)}$ is
obtained from an SDP feasibility program, if we encounter the aforementioned
scenario where $c^{(1)}_{\alpha,\alpha}=0$, we can possibly re-run the
feasibility program with a different starting guess and get a different
$\rho^{(1)}$.  From our numerical simulations, there is good evidence to
suggest that this situation is exceedingly rare.

\end{document}